\documentclass[english,prb, twocolumn, superscriptaddress]{revtex4}
\usepackage[T1]{fontenc}
\usepackage[latin9]{inputenc}
\usepackage{amsmath}
\usepackage{amssymb}
\usepackage{graphicx}
\usepackage{esint,color}

\makeatletter
\@ifundefined{textcolor}{}
{%
 \definecolor{BLACK}{gray}{0}
 \definecolor{WHITE}{gray}{1}
 \definecolor{RED}{rgb}{1,0,0}
 \definecolor{GREEN}{rgb}{0,1,0}
 \definecolor{BLUE}{rgb}{0,0,1}
 \definecolor{CYAN}{cmyk}{1,0,0,0}
 \definecolor{MAGENTA}{cmyk}{0,1,0,0}
 \definecolor{YELLOW}{cmyk}{0,0,1,0}
}

\makeatother

\usepackage{babel}
\begin{document}

\title{Light-Matter Interaction and Lasing in Semiconductor Nanowires:\\ A combined FDTD and Semiconductor Bloch Equation Approach}

\author{Robert Buschlinger}
\affiliation{{\small{Institute of Optics, Information and Photonics, University
of Erlangen-N\"urnberg, 91058 Erlangen, Germany}}}

\author{Michael Lorke}
\affiliation{{\small{\textsuperscript{}Bremen Center for Computational Materials
Science BCCMS, University of Bremen, 28359 Bremen, Germany}}}

\author{Ulf Peschel}
\affiliation{{\small{Institute of Optics, Information and Photonics, University
of Erlangen-N\"urnberg, 91058 Erlangen, Germany}}}

\begin{abstract}
We present a time-domain model for the simulation of light-matter
interaction in semiconductors in arbitrary geometries and across a
wide range of excitation conditions. The electromagnetic field is
treated classically using the finite-difference time-domain method.
The polarization and occupation numbers of the semiconductor material
are described using the semiconductor Bloch equations including many-body effects in the screened
Hartree-Fock approximation. Spontaneous emission noise is introduced
using stochastic driving terms. As an application, we
present simulations of the dynamics of a nanowire laser including optical pumping, seeding by spontaneous emission and the selection of lasing modes.
\end{abstract}


\maketitle



\section{Introduction}

Semiconductor lasers are important components of optical technologies with 
numerous applications in daily life.  Their large scale use, e.g. in optical networks, 
has raised the need for reduced power 
consumption as a contribution to global energy savings.
Therefore, semiconductor nanolasers are at the forefront of  
current research to meet this need.
Recently, semiconductor nanowires have attracted widespread interest
due to their unique combination of electrical and optical properties,
which allow for applications as photonic and plasmonic lasers
or as resonators for the observation of polaritonic effects.\cite{duansingle2003, oultonplasmon2009, Sidiropoulos2014,  saxenaoptically2013, doi:10.1021/nl401355b} 
Many applications rely on sophisticated geometries, where bending and folding of individual wires or the interaction
between multiple wires play an important role in determining the optical properties.\cite{doi:10.1021/nl1040308, 6302326}

On the other hand, the performance of a nanolaser device relies 
heavily on the electronic properties of the respective gain material. In nanowire laser systems, the gain is typically provided by the wire  material itself. 
Alternatively additional constituents such as incorporated transition metal or rare-earth ions can provide the gain.
For semiconductor active materials, the single-particle electronic states are defined by structural 
properties such as material composition, quantum confinement and strain. 

The optical properties, however, are additionally influenced by many-body effects of the excited carriers. 
In the weakly excited regime, these effects give rise to excitonic absorption peaks lying below the band gap.
In the high excitation regime, the excitation dependence of the optical gain is
dominated by phase-space filling and many-body energy renormalizations. Coulomb enhancement of 
interband transitions, screening, and excitation induced dephasing additionally contribute
to magnitude and spectral distribution of the gain \cite{ElSayed:94b,Chow_Koch:99,haug2004quantum,Manzke:02}. 
Therefore, a predictive theory of nanowire lasing needs to incorporate these effects as well as the device geometry.

Numerical methods like finite-difference time-domain\cite{1} (FDTD)
simulations allow for the modelling of electrodynamics in arbitrarily complex
geometries and can be readily extended to solve additional differential
equations describing dispersive, nonlinear or gain media.
However, while this approach has been used to model semiconductor media using incoherently
coupled two-level systems \cite{Huang:06}, 
the combination of the FDTD method with a proper semiconductor gain model,
including excitonic and band-gap renormalization effects is still lacking.
To bridge this gap,
we present a theoretical model combining semiconductor Bloch equations\cite{haug2004quantum} with the FDTD method. 
The Coulomb interaction is treated in the screened Hartree-Fock approximation\cite{Chow_Koch:99} to provide for an accurate description
across different excitation density regimes. 
Since nanowire lasing modes usually have a non-trivial polarization structure, several bands and electronic transitions
need to be included in order to describe the coupling of different electric field polarizations to the material.

As a first application, we present simulations of the dynamics of an optically pumped semiconductor
nanowire, including the amplification of spontaneous emission up to
the onset of lasing and the selection of transverse and longitudinal
modes.




\section{Theoretical Model}

\subsection{General formulation}

To investigate light-matter interaction in semiconductor NWs, we solve
Maxwells equations
\begin{equation}
\frac{\partial}{\partial t}\vec{D}=\frac{1}{\mu_{0}}\nabla\times\vec{B}
\end{equation}

\begin{equation}
\vec{D}=\varepsilon\varepsilon_0\vec{E}+\vec{P}
\end{equation}

\begin{equation}
\frac{\partial}{\partial t}\vec{B}=-\nabla\times\vec{E}
\end{equation}
using the standard FDTD algorithm \cite{15,1} and taking into account the full vectorial character of the electric field $\vec{E}$, the magnetic induction $\vec{B}$ and the dielectric displacement $\vec{D}$. The passive material response is incorporated in a nondispersive dielectric constant $\varepsilon\left(\vec{r}\right)$, while the dynamic action of the semiconductor material is represented by the polarization $\vec{P}$, which  includes contributions from all optical transitions from valence bands with indices $\lambda$ to the conduction band (index $e$)  at electron Bloch vectors $\vec{k}$.
We assume the microscopic polarizations $\psi_{\lambda,q,k}$ as well as the occupation numbers $n_{s,k}$ for conduction-band electrons ($s=e$) and for holes in the different valence bands ($s=\lambda$)  to depend only on the absolute value $k$ of the Bloch vector, so that the polarization in a bulk semiconductor takes the form
\begin{equation}
\vec{P}=2\Re\left(\sum_{\lambda,q}\int_{k}\text{d}k\frac{k^{2}}{\pi^{2}}\vec{d}_{\lambda,q,k}\psi_{\lambda,q,k}\right),
\end{equation}
where $\vec{d}_{\lambda,q,k}$ is the dipole matrix element attributed to the transition from valence band $\lambda$ to the conduction band and coupling to the electric field component pointing in direction $q$. To explicitly take into account the nature of the active material,
we employ semiconductor Bloch equations (SBE) \cite{haug2004quantum} in the screened 
exchange-Coulomb hole approximation \cite{Chow_Koch:99}.

This leads to equations of motion for the microscopic interband polarisations 
$\psi_{\lambda,q,k}$ of the form 
\begin{multline}
i\hbar\frac{\partial}{\partial t}\psi_{\lambda,q,k}=\left(1-n_{e,k}-n_{\lambda,k}\right)\Omega_{\lambda,q,k}\\
+\left(\varepsilon_{e,k}+\varepsilon_{\lambda,k}+\varepsilon_{\lambda,gap}-\triangle\varepsilon_{k}-i\gamma\left(N\right)\right)\psi_{\lambda,q,k}+\Gamma_{\psi,\lambda,q,k},\label{eq:psi}~
\end{multline}
with the renormalized Rabi frequency
\begin{equation}
\Omega_{\lambda,q,k}=\vec{d}_{\lambda,q,k}\vec{E}+\int_{k'}\text{d}k'W_{k,k'}\psi_{\lambda,q,k'},
\end{equation}
using the 
screened Coulomb matrix elements $W_{k,k'}$.
We include an excitation density dependent dephasing $\gamma\left(N\right)$  
caused by carrier-carrier Coulomb interaction \cite{PSSB:PSSB179}.
Transition energies are defined by the band gap energy  $\varepsilon_{\lambda,gap}$ and the
renormalized single-particle energies
\begin{equation}
\varepsilon_{s,k}=\frac{\hbar^{2}k^{2}}{2m_{\text{eff},s}}-\int_{k'}\text{d}k'W_{k,k'}n_{s,k'},
\end{equation}
where $m_{\text{eff},s}$ are the effective masses. The Coulomb hole contribution can be written as
\begin{equation}
\triangle\varepsilon_{k}=\int_{k'}\text{d}k'\left(W_{k,k'}-V_{k,k'}\right)~,
\end{equation}
where  $V_{k,k'}$ is the unscreened 
Coulomb matrix element \cite{Binder1995307}. 
The screened and unscreened Coulomb interaction is given by
\begin{equation}
W_{\left|\vec{k}-\vec{k}'\right|}=\frac{e^{2}}{\varepsilon\varepsilon_{0}}\frac{1}{\left|\vec{k}-\vec{k}'\right|^{2}+\kappa^{2}}
\end{equation} 
with screening wavenumber 
\begin{equation}
\kappa=\sqrt{\frac{e^{2}}{\pi^{2}\varepsilon_{0}\varepsilon_{r}\hbar}\sum_{s\in\left\{ e,\lambda\right\} }m_{eff,s}\int dkn_{s,k}}
\end{equation}
and by
\begin{equation}
V_{\left|\vec{k}-\vec{k}'\right|}=\frac{e^{2}}{\varepsilon\varepsilon_{0}}\frac{1}{\left|\vec{k}-\vec{k}'\right|^{2}}~,
\end{equation}
respectively.
To evaluate these matrix elements numerically, we utilize
an angle-averaging procedure, which is described in \cite{haug2004quantum}.
This yields the matrix elements $V_{k,k'}$ and $W_{k,k'}$ that are used above.

Similar to the equation of motion for the transition amplitudes,
the time evolution of the occupation numbers is given by
\begin{multline}
\frac{\partial}{\partial t}n_{e,k}=-\frac{2}{\hbar}\sum_{\lambda,q}\Im\left(\Omega_{\lambda,q,k}\psi_{\lambda,q,k}^{*}\right)-\gamma_{rec}\sum_{\lambda}n_{\lambda,k}n_{e,k}\\
+\gamma_{f,e}\left(f_{e,k}-n_{e,k}\right)+\Gamma_{n_{e},k}
\label{eq:number}
\end{multline}
for electrons and by
\begin{multline}
\frac{\partial}{\partial t}n_{\lambda,k}=-\frac{2}{\hbar}\sum_{q}\Im\left(\Omega_{\lambda,q,k}\psi_{\lambda,q,k}^{*}\right)-\gamma_{rec}n_{\lambda,k}n_{e,k}\\
+\gamma_{f,h}\left(f_{\lambda,k}-n_{\lambda,k}\right)+\sum_{\lambda'\neq \lambda}\triangle_{\lambda\lambda'k}+\Gamma_{n_{\lambda},k}\label{eq:number2}.
\end{multline}
for holes in the valence band $\lambda$.
The first term describes the carrier excitation by the
electromagnetic field. The two terms involving $\gamma_\text{rec}$
and $\gamma_{f}$ respectively phenomenologically describe non-radiative recombination 
and intra-band relaxation of carriers towards Fermi-Dirac distributions with a band dependent fermi level. $f_{s,k}$ \cite{Huang:06}.
To allow for the relaxation between valence bands, an additional contribution 
$\sum_{\lambda'\neq \lambda}\triangle_{\lambda\lambda'k}\label{eq:sbebesetzung-1-1}$
is included in the equation of motion for the hole populations, where
\begin{equation}
\triangle_{\lambda\lambda'k}=\gamma_{\lambda'\lambda}n_{\lambda',k}(1-n_{\lambda,k})-\gamma_{\lambda\lambda'}n_{\lambda,k}(1-n_{\lambda',k}).\label{eq:delta_intersubband}
\end{equation}

To allow for the description of spontaneous emission in our semiclassical
model, we add noise terms as previously used for two-level systems\cite{PhysRevA.82.063835,Andreasen:09}
\begin{equation}
\Gamma_{\psi,\lambda,q,k}=\left(\xi_{1,\lambda,q,k}+i\xi_{2,\lambda,q,k}\right)\sqrt{\gamma_{se,k}n_{e,k}n_{\lambda,k}}\label{eq:spontem1}
\end{equation}
\[
\gamma_{se,k}=\gamma-\frac{1}{2}\gamma_{rec}
\]
to the right hand side of equation \ref{eq:psi} as well as 
\begin{equation}
\Gamma_{n_{e},k}=\sum_{\lambda}\xi_{3,\lambda,k}\sqrt{n_{e,k}n_{\lambda,k}\gamma_{rec}}
\end{equation}
\begin{equation}
\Gamma_{n_{\lambda},k}=\xi_{3,\lambda,k}\sqrt{n_{e,k}n_{\lambda,k}\gamma_{rec}}.
\end{equation}
to the right hand sides of equations \ref{eq:number} and \ref{eq:number2}. 
$\xi$ are gaussian random numbers with zero mean. The
individual $\psi_{\lambda,q,k}$ are coupled by the Coulomb interaction, therefore excitonic
effects are included in the spontaneous emission, which lies mainly below the band gap. The intra-band relaxation processes associated with $\gamma_{f,e}$ and $\gamma_{f,h}$ are microscopically caused by a combination of carrier-carrier Coulomb scattering and carrier-phonon scattering and redistribute the populations towards quasi-Fermi distributions.
These processes do not describe a decay of polarization and conserve quasi-particle densities. Therefore associated additional noise amplitudes can be safely neglected, at least for high excitation densities and in the lasing regime.

\subsection{Optical transitions for 2-6 Semiconductors}

\begin{figure}
\includegraphics[scale=0.3]{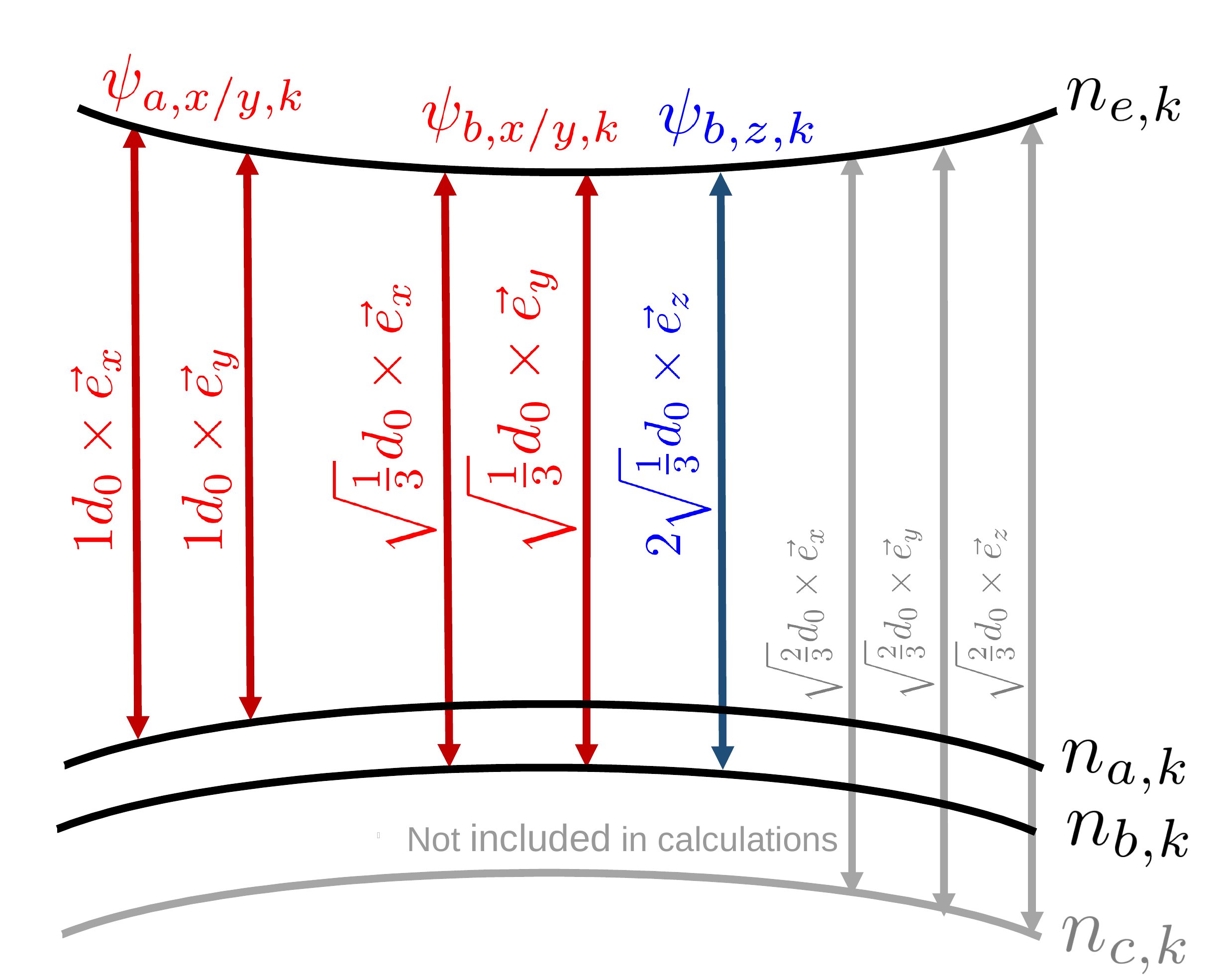}\caption{\label{fig:BandsTwoSix} (Color online). Bands and transitions with their relative dipole matrix elements for the case of a 2-6 semiconductor with wurtzite structure and a crystal axis pointing into the $z$-direction. Greyed out items are not included in the calculations.}
\end{figure}

Simulations of photonic nanostructures require a fully vectorial treatment including the coupling of all electric field components to the material system. Therefore the consideration of all transition elements between bands in a relevant frequency range is necessary. In this section we discuss the involved bands and the possible transitions for the case of 2-6 semiconductors with wurtzite structure (Fig. \ref{fig:BandsTwoSix}).  These materials are most interesting in the case of nanowire lasers. However our algorithm can be easily modified to simulate other band structures. 

Photonic wires with wurtzite structure are birefringent with the optical axes pointing in $z$-direction along the wire. The optical properties arise from transitions between three p-like valence bands (occupation numbers $n_{\lambda,k}$, $\lambda\in\left\{ a,b,c\right\} $) and a single s-like conduction band (occupation number $n_{e,k}$).\cite{PhysRev.116.573,PhysRev.128.2135}

The degeneracy of the valence bands $a$ (total angular momentum $j=\frac{3}{2}$, $z$-component $m_{j}=\pm\frac{3}{2}$),
$b$ ($j=\frac{3}{2},m_{j}=\pm\frac{1}{2}$) and $c$ ($j=\frac{1}{2}$,
$m_{j}=\pm\frac{1}{2}$) is lifted due to spin-orbit coupling and
interaction with the hexagonal crystal field. Valence band $c$ can
be safely neglected in a description of the optical properties, since its energy
 is significantly set off from the fundamental band gap. Thus, the
relevant occupation numbers for our model are $n_{e,k},n_{a,k},n_{b,k}$. 

To find all possible transitions and their respective relative dipole
matrix elements, states are expanded in spin- and orbital angular
momentum eigenstates $|l,m_{l}>|m_{s}>$ using Clebsch-Gordan
coefficients. Assuming conservation of electron spin, allowed optical
transitions are characterized by a conservation of angular momenta of photons and electrons.
Light polarized linearly along the
$z$-axis aligned with the crystals $c$-axis induces transitions with $\Delta m_{l}=0 $.
A transition with $\Delta m_{l}=1$
corresponds to  circular polarization in the
$xy$-plane perpendicular to the $c$-axis. These transitions are expanded in a linear polarization basis to allow for coupling to FDTD simulations.

We obtain two possible transitions between valence band $a$ and the
conduction band coupling to $x$- and $y$-polarized light with dipole
matrix elements $\vec{d}_{a,xy,k}=1\times d_{0,a,k}\vec{e}_{xy}$.
Valence band $b$ turns out to be active  for all three
polarization directions and we obtain dipole matrix elements $\vec{d}_{b,xy,k}=\sqrt{\frac{1}{3}}\times d_{0,b,k}\vec{e}_{xy}$
and $\vec{d}_{b,z,k}=2\sqrt{\frac{1}{3}}\times d_{0,b,k}\vec{e}_{z}$.

\subsection{Parameters for CdS}
In this work we focus on CdS nanowire structures. Gap energies for the two
relevant bands $\varepsilon_{gap,a}=2420meV$ and $\varepsilon_{gap,b}=2435meV$
are taken from theoretical calculations.\cite{doi:10.1139/P11-023,PhysRevB.50.10780}
Anisotropic effective masses found in literature\cite{cdsmefflandoltboernstein}
have to be angle-averaged to be applied in our model, where only the absolute
values of the $k$-vector are used. We obtain $m_{\text{eff},e}=0,1619m_{e}$
for electrons as well as $m_{\text{eff},a}=0,5951m_{e}$ and $m_{\text{eff},b}=0,713m_{e}$
for holes. 
The background refractive index $n_{\text{bg}}=\sqrt{\varepsilon}=2.74$ has been
chosen so that reported values for the exciton binding energy\cite{Gutowski2009}
are reproduced. The dipole matrix element $d_0=0.279e\cdot nm$  
is taken from literature.\cite{Hassan199380}

We use a density-dependent dephasing time-constant
similiar to experimental and theoretical literature\cite{Hugel2000,PSSB:PSSB179}

\begin{equation}
\gamma(N)=\gamma_{0}+\gamma_{a}N_{e}^{0.3},
\end{equation}

with $\gamma_{0}=5ps^{-1},\gamma_{a}=4\times10^{-5}\frac{cm}{ps}$. 
Typical values for relaxation and recombination times are $\gamma_{rec}=10^{9}s^{-1}$,
$\gamma_{f,e}=10^{12}s^{-1}$ and $\gamma_{f,h}=10^{13}s^{-1}$. Holes
relax from valence band $b$ to valence band $a$, so that Eq.
\ref{eq:delta_intersubband} becomes
\begin{equation}
\triangle_{ab,k}=\gamma_{ba}n_{b,k}(1-n_{a,k}),
\end{equation}
\begin{equation}
\triangle_{ba,k}=-\triangle_{ab,k}
\end{equation}
using the relaxation rate\cite{Qi1988575} $\gamma_{ba}=6\times10^{9}s^{-1}$.

\section{Results}

\begin{figure}
\includegraphics[scale=0.55]{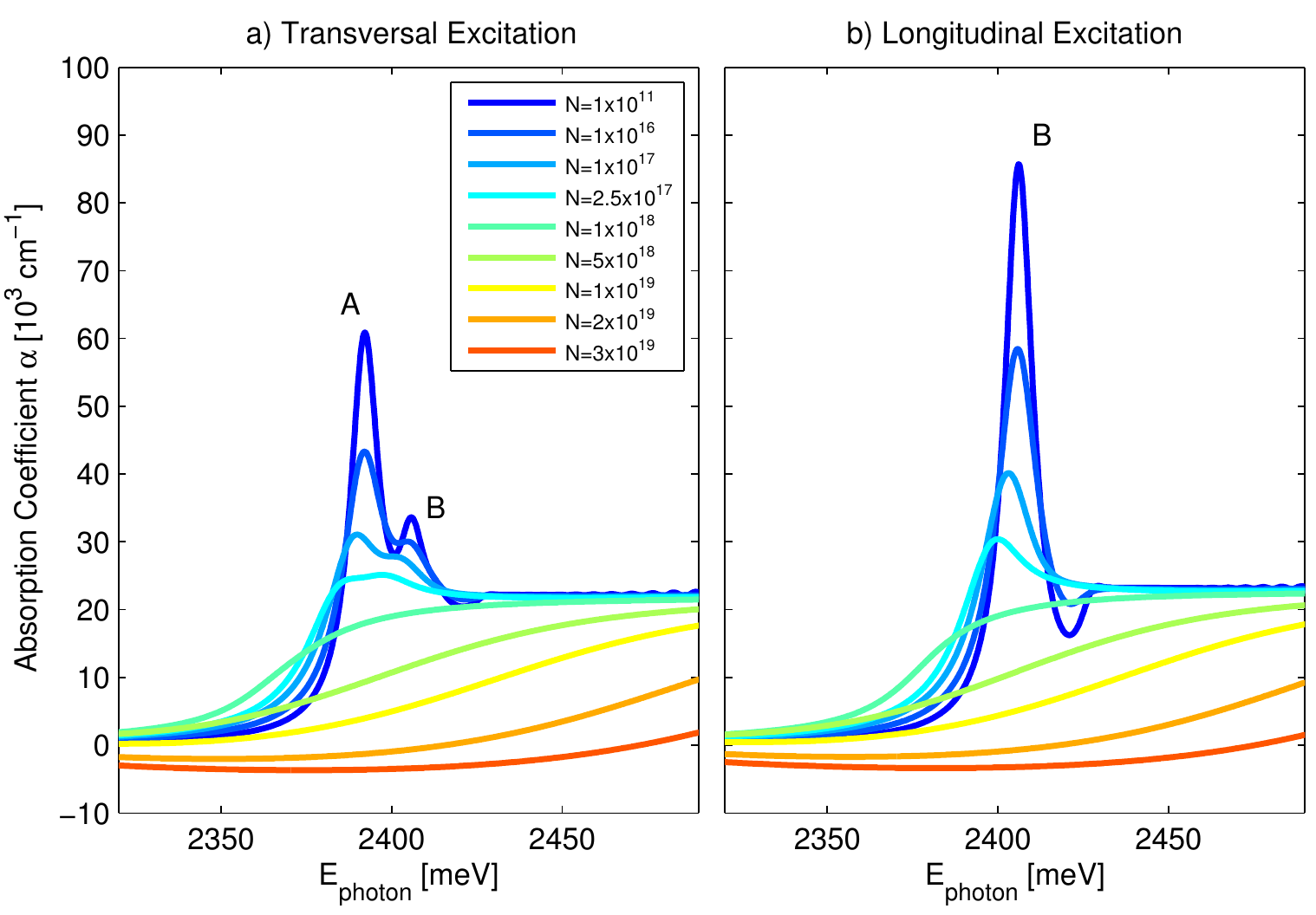}\caption{\label{fig:Linear-absorption-spectra} (Color online). Absorption spectra of
CdS  calculated with the multiband semiconductor Bloch equations model
for different excitation densities $N$. a) Excitation polarized perpendicular
to the crystals $c$-axis. b) Excitation polarized along
the crystals $c$-axis. The peaks labelled ``A'' and ``B''
correspond to the exciton 1s-resonances of the transitions respectively
involving valence band $a$ and valence band $b$.}
\end{figure}

In order to probe the linear response of our material model, we simulate
a thin CdS sample in air irradiated with a plane wave and record
the spectrum of the electric field $\vec{E}\left(\omega\right)$ and the polarization $\vec{P}\left(\omega\right)$ inside
the sample. From this, the electronic contribution to the permittivity $\delta\chi_{ij}=\frac{P_i}{\varepsilon_{0}E_j}$
is calculated and the absorption coefficient $\alpha$ is obtained.
For an excitation polarized perpendicularly to the crystals $c$-axis
($E_{x/y}$), the absorption spectrum shows two distinct peaks corresponding
to the exciton 1s resonances of the transitions between each of the
valence bands and the conduction band (Fig. \ref{fig:Linear-absorption-spectra}(a)).
For an excitation along the $c$-axis ($E_{z}$), only one absorption
peak can be observed, since for this polarization only the transition
from valence band $b$ to the conduction band is accessible (Fig.
\ref{fig:Linear-absorption-spectra}(b)). In both cases the exciton
peaks vanish for higher excitation densities due to the combined effects
of screening and band filling. 

\begin{figure}
\includegraphics[scale=1]{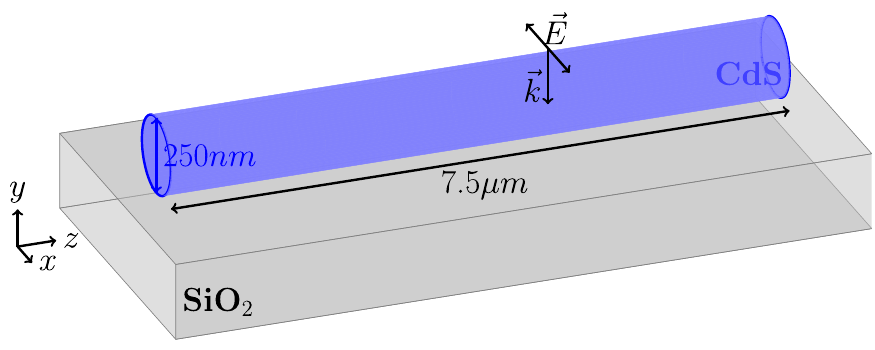}
\caption{\label{fig:sketch} (Color online). Sketch of the simulated nanowire geometry including polarization and propagation direction of the exciting pulse. The wire is centered on $z=0$}
\end{figure}

\begin{figure}
\includegraphics[scale=0.55]{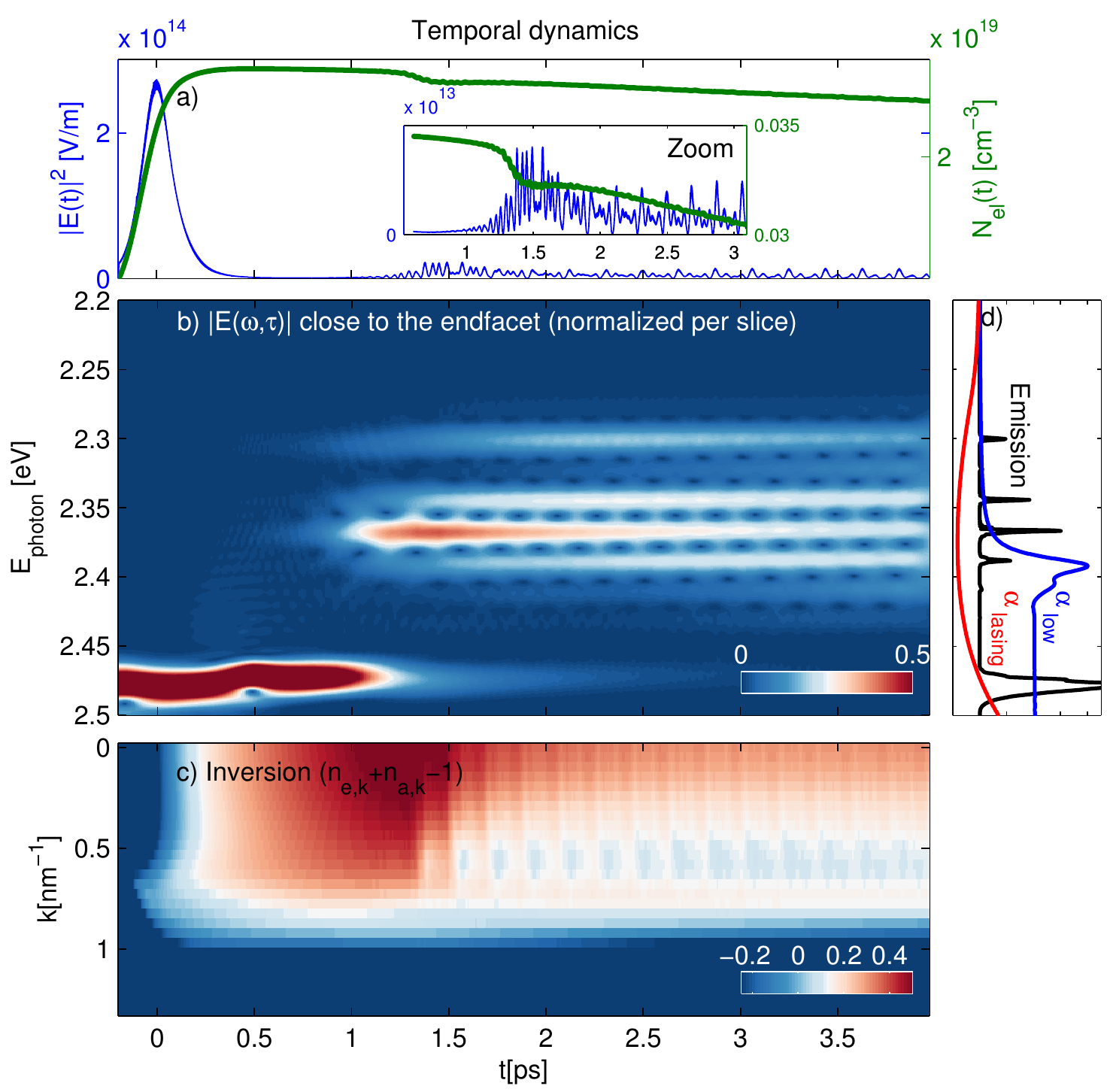}\caption{\label{fig:TempDynamicsEx} (Color online). Temporal field dynamics inside a nanowire
laser ($l=7.5\mu m$, $d=250nm$) excited with a pump pulse (central wavelength $\lambda_0=500nm$, temporal width $w_t=100fs$) polarized perpendicularly to the wire axis. a)
Dynamics of electron volume density and electric field intensity inside
the wire. The inset shows a zoom into the temporal region of the initial
laser pulse. b) Temporal dynamics of the individual longitudinal modes
obtained from a windowed fourier transform of the temporal fields (Normalization per timeslice). 
c) Population dynamics in $k$-space. d) Spectrum showing
the emission lines (black). For better orientation, the absorption coefficients for
a low excitation density (blue) and a high excitation density (red)
similiar to the densities present after the absorption of the pump
pulse are included. }
\end{figure}

\begin{figure}
\includegraphics[scale=0.55]{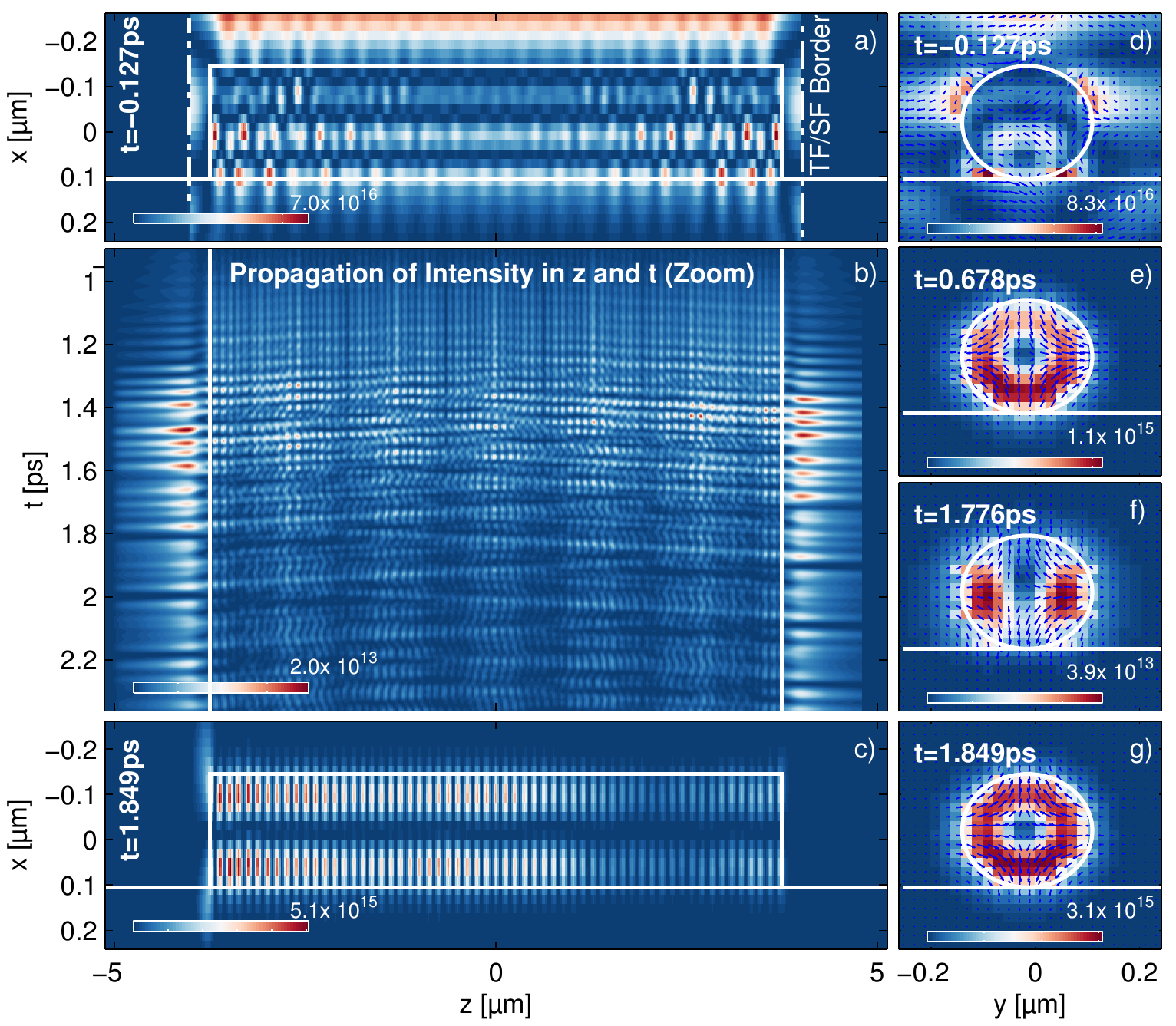}\caption{\label{fig:SpatTempDynamicsEx} (Color online). Spatiotemporal field dynamics inside
a nanowire laser($l=7.5\mu m$, $d=250nm$). (a),(d): Electric field intensity in an $xz$- and an $xy$-slice of the simulation volume during pumping. (b): $|E|^2$ averaged across transverse slices and plotted along $z$ and $t$. (e-g): Transverse intensity profiles inside the wire during lasing emission. (c): Intensity profile in an $xz$-slice after the maximum of the lasing emission.}
\end{figure}

We now turn our attention to the lasing dynamics of optically excited
nanowire lasers. We simulate a nanowire with diameter $d=250nm$ and length $l=7.5\mu m$ extending along the $z$-axis centered on $z=0$ and resting on a fused silica substrate as shown in Fig. \ref{fig:sketch}. The wire is pumped from above ($y$-direction) with a $x$-polarized plane wave pump pulse with a sech-shaped time dependence (central wavelength $\lambda_0=500nm$, temporal width  $w_{t}=100fs$).  To inject this excitation pulse, we utilize the total field / scattered field (TF/SF) formalism,\cite{1} which allows for the excitation of a plane wave inside a bounded simulation volume. Outside the TF/SF border, only fields scattered or emitted by the nanowire are present.
Fig. \ref{fig:TempDynamicsEx}(a) shows the temporal dynamics of the electric field and of the electron density. The electron density is recorded in a slice perpendicular to the wire axis positioned close to an endfacet ($z=3.6µm$). The electric field intensity is averaged across a slice outside the wire ($z=4.2µm$). After applying a windowed Fourier transform to the time-domain data, the dynamics of the involved modes can be studied (Fig. \ref{fig:TempDynamicsEx}(b)).

Starting from thermal equilibrium, the conduction band electron density is pumped by the excitation pulse, which is centered at $t=0$. After the passage of the excitation pulse, intensities drop until stimulated emission sets in with a steep rise at about $t=1ps$ (Inset of Fig. \ref{fig:TempDynamicsEx}(a)). 
The presence of equally spaced modes in the time-dependent spectra (Fig. \ref{fig:TempDynamicsEx}(b)) indicates lasing emission from approximately $t=1ps$ up to the end of the simulation window. Initially, the emission is dominated by a single longitudinal mode, which is rapidly amplified and depletes the material gain, until the emission reaches its maximum value at $t=1.5ps$. At this point, power is redistributed to other longitudinal modes which can access the remaining material gain. The lasing emission continues with a slowly falling slope, leading to a strongly asymmetric shape of the emitted pulse sequence. Due to interference between the lasing modes the emitted pulse sequence has a rather irregular temporal shape. 

The $k$-resolved inversion of the transition from valence band $a$ to the conduction band is plotted in Fig. \ref{fig:TempDynamicsEx}(c). 
The pump pulse is positioned at higher energies than the plotted Bloch vector states.
The lasing pulse initially depletes excitations in a region of wavevector space around $k=0.6nm^{-1}$. 
This explains, why modes which mostly access other regions of $k$-space are still weakly amplified and continue to lase after the maximum of the overall emission. 

Fig. \ref{fig:SpatTempDynamicsEx}(b) shows the spatiotemporal dynamics of the lasing process. 
The electric field intensity averaged over each $xy$-slice is plotted over time and over the length of the simulation volume along the $z$-direction. Initially, the field maxima inside the wire keep a fixed location along the $z$-axis, as expected for nearly single-mode lasing action. As the initial emission peak is reached, the field profile along the wire gets strongly modulated, since the contribution of additional longitudinal modes becomes relevant.

Apart from longitudinal modes, the properties of a nanowire laser are also strongly influenced by transverse modes. The transverse field structure determines the modal gain as well as the mode reflectivity at the end facets of the nanowire.\cite{:/content/aip/journal/apl/83/6/10.1063/1.1599037,1337019}
Figs. \ref{fig:SpatTempDynamicsEx}(d-g) show the transverse fields across a slice ($z=0$) of the nanowire at different times during the simulation. Figs. \ref{fig:SpatTempDynamicsEx}(a,d) display the scattering of the incident pump light on the nanowire, leading to a spatially inhomogeneous pump profile. In Figs. \ref{fig:SpatTempDynamicsEx}(e-g), the mode dynamics after the onset of lasing can be observed. As  predicted from linear calculations,\cite{Roeder2014} the field profiles of wires of the investigated diameter range are dominated by the HE21 mode, which has a high modal reflectivity as well as a high confinement factor. However, the field profile is fluctuating between the individual timesteps, indicating an admixture of additional transverse modes. This is especially visible in panel (f), where the polarization of the HE21 mode is still recognizable, but the intensity profile differs significantly from that of the pure mode. Panels (e) and (g) show almost pure HE21 mode profiles.
\begin{figure}
\includegraphics[scale=0.55]{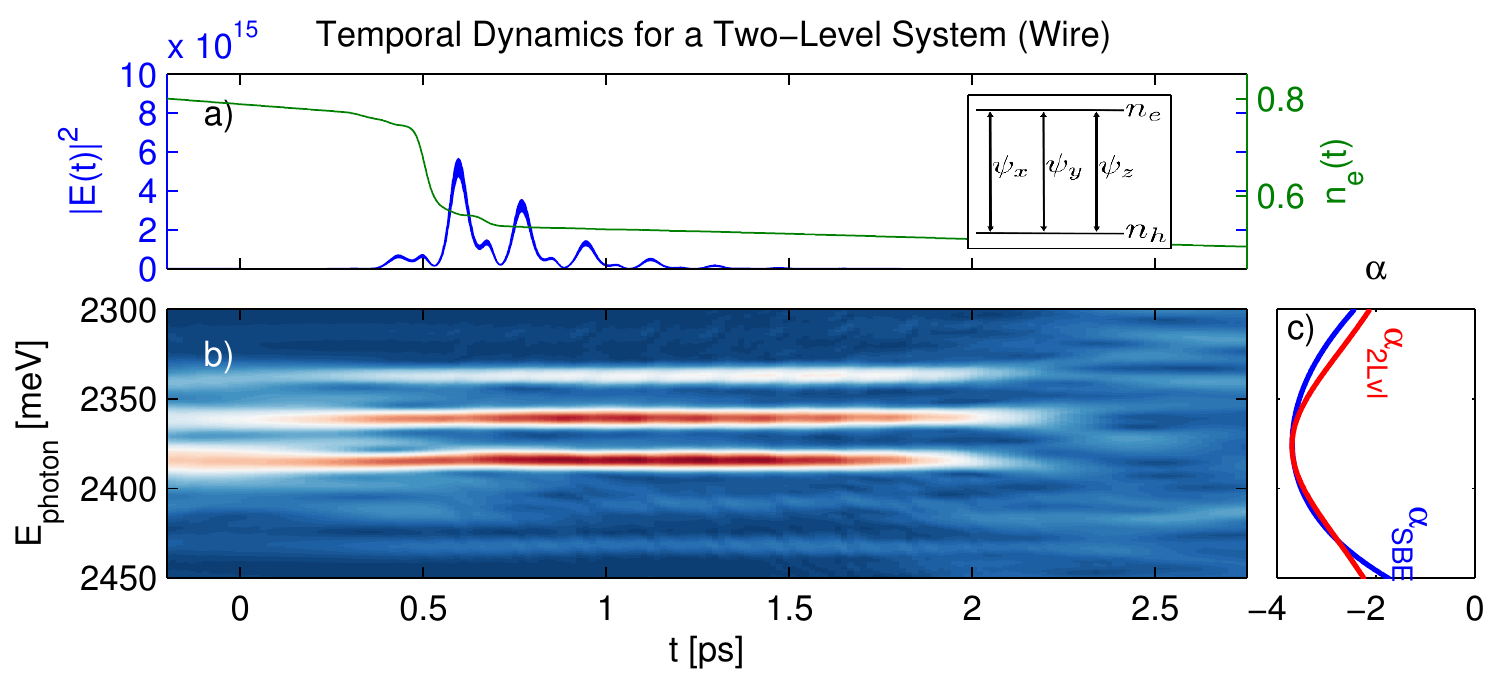}\caption{\label{fig:SpatTempDynamics2Lvl} (Color online). Dynamics inside
a nanowire laser($l=7.5\mu m$, $d=250nm$) consisting of a two-level material with a gain profile fitted to that of highly excited CdS (See panel(c)). (a): Temporal dynamics of field strength $|E|^2$ and excitation density $n_e$. (b): Mode dynamics obtained from windowed fourier transform (normalized per timeslice). (c): Absorption spectrum of the two-level system (red). The absorption spectrum for the full model is plotted in blue for comparison.}
\end{figure}

As the numerical treatment of semiconductor Bloch equations is quite cumbersome one might be tempted to restrict to simpler material models. Indeed some of the observed dynamical features are not inherent to the semiconductor material, but more to the wire geometry. 
However others require the material model to be reproduced.
For comparison, we include a simulation of a nanowire laser where the active material is described by a two-level system, as often employed in FDTD simulations. The transition between the two levels couples equally to all three field polarizations (see Inset of Fig.\ref{fig:SpatTempDynamics2Lvl}(a) ). The material is assumed to be excited to an upper-level occupation of $n_e=0.8$ and material parameters are tuned to fit the gain profile of highly excited CdS. 
We observe, that even though the gain profile (Fig. \ref{fig:SpatTempDynamics2Lvl}(c)) looks very similiar to that of the semiconductor material, the resulting lasing dynamics differ considerably. As opposed to the full model, there is no prolonged stimulated emission after the initial peak, leading to a more symmetric temporal shape of the emission. (Fig. \ref{fig:SpatTempDynamics2Lvl}(a)).
Since the shape of the gain profile of a single two-level system does not change as carriers are depleted, the spectral shape of the emission stays approximately constant during lasing action (Fig. \ref{fig:SpatTempDynamics2Lvl}(b)). Lasing emission stops as soon as the dominant mode no longer fulfills the lasing condition.

\section{Conclusions}
We presented a theoretical model for the simulation of light-matter interaction in arbitrarily shaped semiconductor structures based on the FDTD method and the semiconductor Bloch equations. We adapted the model to the band structure of 2-6 semiconductors and presented simulations of the lasing dynamics of an optically pumped CdS-nanowire. Our model allows for the description of semiconductors across the whole range of excitation conditions from the weakly excited case to the lasing regime and is not limited to equilibrium distributions of carriers either in Bloch vector or real space. Thus it is especially interesting for lasing simulations where excitation conditions vary strongly across space or time as well as for simulations of nonlinear optical phenomena in the weakly to moderately excited regime.

\begin{acknowledgments}
The authors gratefully acknowledge financial support by Deutsche Forschungsgemeinschaft
(Forschergruppe FOR1616, projects P5 and E4). Robert Buschlinger acknowledges financial support from the International
Max Planck Research School Physics of Light.
\end{acknowledgments}


\begin{thebibliography}{30}
\expandafter\ifx\csname natexlab\endcsname\relax\def\natexlab#1{#1}\fi
\expandafter\ifx\csname bibnamefont\endcsname\relax
  \def\bibnamefont#1{#1}\fi
\expandafter\ifx\csname bibfnamefont\endcsname\relax
  \def\bibfnamefont#1{#1}\fi
\expandafter\ifx\csname citenamefont\endcsname\relax
  \def\citenamefont#1{#1}\fi
\expandafter\ifx\csname url\endcsname\relax
  \def\url#1{\texttt{#1}}\fi
\expandafter\ifx\csname urlprefix\endcsname\relax\def\urlprefix{URL }\fi
\providecommand{\bibinfo}[2]{#2}
\providecommand{\eprint}[2][]{\url{#2}}

\bibitem[{\citenamefont{Duan et~al.}(2003)\citenamefont{Duan, Huang, Agarwal,
  and Lieber}}]{duansingle2003}
\bibinfo{author}{\bibfnamefont{X.}~\bibnamefont{Duan}},
  \bibinfo{author}{\bibfnamefont{Y.}~\bibnamefont{Huang}},
  \bibinfo{author}{\bibfnamefont{R.}~\bibnamefont{Agarwal}}, \bibnamefont{and}
  \bibinfo{author}{\bibfnamefont{C.~M.} \bibnamefont{Lieber}},
  \bibinfo{journal}{Nature} \textbf{\bibinfo{volume}{421}},
  \bibinfo{pages}{241} (\bibinfo{year}{2003}),
  \urlprefix\url{http://dx.doi.org/10.1038/nature01353}.

\bibitem[{\citenamefont{Oulton et~al.}(2009)\citenamefont{Oulton, Sorger,
  Zentgraf, Ma, Gladden, Dai, Bartal, and Zhang}}]{oultonplasmon2009}
\bibinfo{author}{\bibfnamefont{R.~F.} \bibnamefont{Oulton}},
  \bibinfo{author}{\bibfnamefont{V.~J.} \bibnamefont{Sorger}},
  \bibinfo{author}{\bibfnamefont{T.}~\bibnamefont{Zentgraf}},
  \bibinfo{author}{\bibfnamefont{R.-M.} \bibnamefont{Ma}},
  \bibinfo{author}{\bibfnamefont{C.}~\bibnamefont{Gladden}},
  \bibinfo{author}{\bibfnamefont{L.}~\bibnamefont{Dai}},
  \bibinfo{author}{\bibfnamefont{G.}~\bibnamefont{Bartal}}, \bibnamefont{and}
  \bibinfo{author}{\bibfnamefont{X.}~\bibnamefont{Zhang}},
  \bibinfo{journal}{Nature} \textbf{\bibinfo{volume}{461}},
  \bibinfo{pages}{629} (\bibinfo{year}{2009}),
  \urlprefix\url{http://dx.doi.org/10.1038/nature08364}.

\bibitem[{\citenamefont{Sidiropoulos et~al.}(2014)\citenamefont{Sidiropoulos,
  Roder, Geburt, Hess, Maier, Ronning, and Oulton}}]{Sidiropoulos2014}
\bibinfo{author}{\bibfnamefont{T.~P.~H.} \bibnamefont{Sidiropoulos}},
  \bibinfo{author}{\bibfnamefont{R.}~\bibnamefont{Roder}},
  \bibinfo{author}{\bibfnamefont{S.}~\bibnamefont{Geburt}},
  \bibinfo{author}{\bibfnamefont{O.}~\bibnamefont{Hess}},
  \bibinfo{author}{\bibfnamefont{S.~A.} \bibnamefont{Maier}},
  \bibinfo{author}{\bibfnamefont{C.}~\bibnamefont{Ronning}}, \bibnamefont{and}
  \bibinfo{author}{\bibfnamefont{R.~F.} \bibnamefont{Oulton}},
  \bibinfo{journal}{Nat Phys} \textbf{\bibinfo{volume}{advance online
  publication}} (\bibinfo{year}{2014}), ISSN \bibinfo{issn}{1745-2481},
  \urlprefix\url{http://dx.doi.org/10.1038/nphys3103 10.1038/nphys3103
  http://www.nature.com/nphys/journal/vaop/ncurrent/abs/nphys3103.html\#supplementary-information}.

\bibitem[{\citenamefont{Saxena et~al.}(2013)\citenamefont{Saxena, Mokkapati,
  Parkinson, Jiang, Gao, Tan, and Jagadish}}]{saxenaoptically2013}
\bibinfo{author}{\bibfnamefont{D.}~\bibnamefont{Saxena}},
  \bibinfo{author}{\bibfnamefont{S.}~\bibnamefont{Mokkapati}},
  \bibinfo{author}{\bibfnamefont{P.}~\bibnamefont{Parkinson}},
  \bibinfo{author}{\bibfnamefont{N.}~\bibnamefont{Jiang}},
  \bibinfo{author}{\bibfnamefont{Q.}~\bibnamefont{Gao}},
  \bibinfo{author}{\bibfnamefont{H.~H.} \bibnamefont{Tan}}, \bibnamefont{and}
  \bibinfo{author}{\bibfnamefont{C.}~\bibnamefont{Jagadish}},
  \bibinfo{journal}{Nature Photonics} \textbf{\bibinfo{volume}{7}},
  \bibinfo{pages}{963} (\bibinfo{year}{2013}),
  \urlprefix\url{http://dx.doi.org/10.1038/nphoton.2013.303}.

\bibitem[{\citenamefont{Roeder et~al.}(2013)\citenamefont{Roeder, Wille,
  Geburt, Rensberg, Zhang, Lu, Capasso, Buschlinger, Peschel, and
  Ronning}}]{doi:10.1021/nl401355b}
\bibinfo{author}{\bibfnamefont{R.}~\bibnamefont{Roeder}},
  \bibinfo{author}{\bibfnamefont{M.}~\bibnamefont{Wille}},
  \bibinfo{author}{\bibfnamefont{S.}~\bibnamefont{Geburt}},
  \bibinfo{author}{\bibfnamefont{J.}~\bibnamefont{Rensberg}},
  \bibinfo{author}{\bibfnamefont{M.}~\bibnamefont{Zhang}},
  \bibinfo{author}{\bibfnamefont{J.~G.} \bibnamefont{Lu}},
  \bibinfo{author}{\bibfnamefont{F.}~\bibnamefont{Capasso}},
  \bibinfo{author}{\bibfnamefont{R.}~\bibnamefont{Buschlinger}},
  \bibinfo{author}{\bibfnamefont{U.}~\bibnamefont{Peschel}}, \bibnamefont{and}
  \bibinfo{author}{\bibfnamefont{C.}~\bibnamefont{Ronning}},
  \bibinfo{journal}{Nano Letters} \textbf{\bibinfo{volume}{13}},
  \bibinfo{pages}{3602} (\bibinfo{year}{2013}),
  \eprint{http://pubs.acs.org/doi/pdf/10.1021/nl401355b},
  \urlprefix\url{http://pubs.acs.org/doi/abs/10.1021/nl401355b}.

\bibitem[{\citenamefont{Xiao et~al.}(2011)\citenamefont{Xiao, Meng, Wang, Ye,
  Yu, Wang, Gu, Dai, and Tong}}]{doi:10.1021/nl1040308}
\bibinfo{author}{\bibfnamefont{Y.}~\bibnamefont{Xiao}},
  \bibinfo{author}{\bibfnamefont{C.}~\bibnamefont{Meng}},
  \bibinfo{author}{\bibfnamefont{P.}~\bibnamefont{Wang}},
  \bibinfo{author}{\bibfnamefont{Y.}~\bibnamefont{Ye}},
  \bibinfo{author}{\bibfnamefont{H.}~\bibnamefont{Yu}},
  \bibinfo{author}{\bibfnamefont{S.}~\bibnamefont{Wang}},
  \bibinfo{author}{\bibfnamefont{F.}~\bibnamefont{Gu}},
  \bibinfo{author}{\bibfnamefont{L.}~\bibnamefont{Dai}}, \bibnamefont{and}
  \bibinfo{author}{\bibfnamefont{L.}~\bibnamefont{Tong}},
  \bibinfo{journal}{Nano Letters} \textbf{\bibinfo{volume}{11}},
  \bibinfo{pages}{1122} (\bibinfo{year}{2011}),
  \eprint{http://pubs.acs.org/doi/pdf/10.1021/nl1040308},
  \urlprefix\url{http://pubs.acs.org/doi/abs/10.1021/nl1040308}.

\bibitem[{\citenamefont{Xu et~al.}(2012)\citenamefont{Xu, Wright, Luk, Figiel,
  Cross, Lester, Balakrishnan, Wang, Brener, and Li}}]{6302326}
\bibinfo{author}{\bibfnamefont{H.}~\bibnamefont{Xu}},
  \bibinfo{author}{\bibfnamefont{J.~B.} \bibnamefont{Wright}},
  \bibinfo{author}{\bibfnamefont{T.-S.} \bibnamefont{Luk}},
  \bibinfo{author}{\bibfnamefont{J.~J.} \bibnamefont{Figiel}},
  \bibinfo{author}{\bibfnamefont{K.}~\bibnamefont{Cross}},
  \bibinfo{author}{\bibfnamefont{L.}~\bibnamefont{Lester}},
  \bibinfo{author}{\bibfnamefont{G.}~\bibnamefont{Balakrishnan}},
  \bibinfo{author}{\bibfnamefont{G.~T.} \bibnamefont{Wang}},
  \bibinfo{author}{\bibfnamefont{I.}~\bibnamefont{Brener}}, \bibnamefont{and}
  \bibinfo{author}{\bibfnamefont{Q.}~\bibnamefont{Li}},
  \bibinfo{journal}{Applied Physics Letters} \textbf{\bibinfo{volume}{101}},
  \bibinfo{pages}{113106} (\bibinfo{year}{2012}), ISSN
  \bibinfo{issn}{0003-6951}.

\bibitem[{\citenamefont{ElSayed et~al.}(1994)\citenamefont{ElSayed, B{\'a}nyai,
  and Haug}}]{ElSayed:94b}
\bibinfo{author}{\bibfnamefont{K.}~\bibnamefont{ElSayed}},
  \bibinfo{author}{\bibfnamefont{L.}~\bibnamefont{B{\'a}nyai}},
  \bibnamefont{and} \bibinfo{author}{\bibfnamefont{H.}~\bibnamefont{Haug}},
  \bibinfo{journal}{Phys. Rev. B} \textbf{\bibinfo{volume}{\textbf{50}}},
  \bibinfo{pages}{1541} (\bibinfo{year}{1994}).

\bibitem[{\citenamefont{Chow and Koch}(1999)}]{Chow_Koch:99}
\bibinfo{author}{\bibfnamefont{W.}~\bibnamefont{Chow}} \bibnamefont{and}
  \bibinfo{author}{\bibfnamefont{S.}~\bibnamefont{Koch}},
  \emph{\bibinfo{title}{Semiconductor-Laser Fundamentals}}
  (\bibinfo{publisher}{Springer-Verlag}, \bibinfo{address}{Berlin},
  \bibinfo{year}{1999}), \bibinfo{edition}{1st} ed.

\bibitem[{\citenamefont{Haug and Koch}(2004)}]{haug2004quantum}
\bibinfo{author}{\bibfnamefont{H.}~\bibnamefont{Haug}} \bibnamefont{and}
  \bibinfo{author}{\bibfnamefont{S.}~\bibnamefont{Koch}},
  \emph{\bibinfo{title}{Quantum Theory of the Optical and Electronic Properties
  of Semiconductors (4th Edition)}} (\bibinfo{publisher}{World Scientific},
  \bibinfo{year}{2004}), ISBN \bibinfo{isbn}{9789812387561},
  \urlprefix\url{http://books.google.de/books?id=-UoG0Hx0w04C}.

\bibitem[{\citenamefont{Manzke and Henneberger}(2002)}]{Manzke:02}
\bibinfo{author}{\bibfnamefont{G.}~\bibnamefont{Manzke}} \bibnamefont{and}
  \bibinfo{author}{\bibfnamefont{K.}~\bibnamefont{Henneberger}},
  \bibinfo{journal}{phys. stat. sol. (b)}
  \textbf{\bibinfo{volume}{\textbf{234}}}, \bibinfo{pages}{233}
  (\bibinfo{year}{2002}).

\bibitem[{\citenamefont{Taflove and Hagness}(2005)}]{1}
\bibinfo{author}{\bibfnamefont{A.}~\bibnamefont{Taflove}} \bibnamefont{and}
  \bibinfo{author}{\bibfnamefont{S.~C.} \bibnamefont{Hagness}},
  \emph{\bibinfo{title}{Computational Electrodynamics: The Finite-Difference
  Time-Domain Method}} (\bibinfo{publisher}{Artech House},
  \bibinfo{year}{2005}).

\bibitem[{\citenamefont{Huang and Ho}(2006)}]{Huang:06}
\bibinfo{author}{\bibfnamefont{Y.}~\bibnamefont{Huang}} \bibnamefont{and}
  \bibinfo{author}{\bibfnamefont{S.-T.} \bibnamefont{Ho}},
  \bibinfo{journal}{Opt. Express} \textbf{\bibinfo{volume}{14}},
  \bibinfo{pages}{3569} (\bibinfo{year}{2006}),
  \urlprefix\url{http://www.opticsexpress.org/abstract.cfm?URI=oe-14-8-3569}.

\bibitem[{\citenamefont{Yee}(1966)}]{15}
\bibinfo{author}{\bibfnamefont{K.~S.} \bibnamefont{Yee}},
  \bibinfo{journal}{IEEE Transactions on Antennas and Propagation}
  \textbf{\bibinfo{volume}{AP-14}}, \bibinfo{pages}{302}
  (\bibinfo{year}{1966}).

\bibitem[{\citenamefont{Haug}(2000)}]{PSSB:PSSB179}
\bibinfo{author}{\bibfnamefont{H.}~\bibnamefont{Haug}},
  \bibinfo{journal}{physica status solidi (b)} \textbf{\bibinfo{volume}{221}},
  \bibinfo{pages}{179} (\bibinfo{year}{2000}), ISSN \bibinfo{issn}{1521-3951},
  \urlprefix\url{http://dx.doi.org/10.1002/1521-3951(200009)221:1<179::AID-PSSB179>3.0.CO;2-6}.

\bibitem[{\citenamefont{Binder and Koch}(1995)}]{Binder1995307}
\bibinfo{author}{\bibfnamefont{R.}~\bibnamefont{Binder}} \bibnamefont{and}
  \bibinfo{author}{\bibfnamefont{S.}~\bibnamefont{Koch}},
  \bibinfo{journal}{Progress in Quantum Electronics}
  \textbf{\bibinfo{volume}{19}}, \bibinfo{pages}{307 } (\bibinfo{year}{1995}),
  ISSN \bibinfo{issn}{0079-6727},
  \urlprefix\url{http://www.sciencedirect.com/science/article/pii/007967279500001S}.

\bibitem[{\citenamefont{Andreasen and Cao}(2010)}]{PhysRevA.82.063835}
\bibinfo{author}{\bibfnamefont{J.}~\bibnamefont{Andreasen}} \bibnamefont{and}
  \bibinfo{author}{\bibfnamefont{H.}~\bibnamefont{Cao}},
  \bibinfo{journal}{Phys. Rev. A} \textbf{\bibinfo{volume}{82}},
  \bibinfo{pages}{063835} (\bibinfo{year}{2010}),
  \urlprefix\url{http://link.aps.org/doi/10.1103/PhysRevA.82.063835}.

\bibitem[{\citenamefont{Andreasen and Cao}(2009)}]{Andreasen:09}
\bibinfo{author}{\bibfnamefont{J.}~\bibnamefont{Andreasen}} \bibnamefont{and}
  \bibinfo{author}{\bibfnamefont{H.}~\bibnamefont{Cao}}, \bibinfo{journal}{J.
  Lightwave Technol.} \textbf{\bibinfo{volume}{27}}, \bibinfo{pages}{4530}
  (\bibinfo{year}{2009}),
  \urlprefix\url{http://jlt.osa.org/abstract.cfm?URI=jlt-27-20-4530}.

\bibitem[{\citenamefont{Thomas and Hopfield}(1959)}]{PhysRev.116.573}
\bibinfo{author}{\bibfnamefont{D.~G.} \bibnamefont{Thomas}} \bibnamefont{and}
  \bibinfo{author}{\bibfnamefont{J.~J.} \bibnamefont{Hopfield}},
  \bibinfo{journal}{Phys. Rev.} \textbf{\bibinfo{volume}{116}},
  \bibinfo{pages}{573} (\bibinfo{year}{1959}),
  \urlprefix\url{http://link.aps.org/doi/10.1103/PhysRev.116.573}.

\bibitem[{\citenamefont{Thomas and Hopfield}(1962)}]{PhysRev.128.2135}
\bibinfo{author}{\bibfnamefont{D.~G.} \bibnamefont{Thomas}} \bibnamefont{and}
  \bibinfo{author}{\bibfnamefont{J.~J.} \bibnamefont{Hopfield}},
  \bibinfo{journal}{Phys. Rev.} \textbf{\bibinfo{volume}{128}},
  \bibinfo{pages}{2135} (\bibinfo{year}{1962}),
  \urlprefix\url{http://link.aps.org/doi/10.1103/PhysRev.128.2135}.

\bibitem[{\citenamefont{Ekuma et~al.}(2011)\citenamefont{Ekuma, Franklin, Zhao,
  Wang, and Bagayoko}}]{doi:10.1139/P11-023}
\bibinfo{author}{\bibfnamefont{E.~C.} \bibnamefont{Ekuma}},
  \bibinfo{author}{\bibfnamefont{L.}~\bibnamefont{Franklin}},
  \bibinfo{author}{\bibfnamefont{G.~L.} \bibnamefont{Zhao}},
  \bibinfo{author}{\bibfnamefont{J.~T.} \bibnamefont{Wang}}, \bibnamefont{and}
  \bibinfo{author}{\bibfnamefont{D.}~\bibnamefont{Bagayoko}},
  \bibinfo{journal}{Canadian Journal of Physics} \textbf{\bibinfo{volume}{89}},
  \bibinfo{pages}{319} (\bibinfo{year}{2011}),
  \eprint{http://www.nrcresearchpress.com/doi/pdf/10.1139/P11-023},
  \urlprefix\url{http://www.nrcresearchpress.com/doi/abs/10.1139/P11-023}.

\bibitem[{\citenamefont{Zakharov et~al.}(1994)\citenamefont{Zakharov, Rubio,
  Blase, Cohen, and Louie}}]{PhysRevB.50.10780}
\bibinfo{author}{\bibfnamefont{O.}~\bibnamefont{Zakharov}},
  \bibinfo{author}{\bibfnamefont{A.}~\bibnamefont{Rubio}},
  \bibinfo{author}{\bibfnamefont{X.}~\bibnamefont{Blase}},
  \bibinfo{author}{\bibfnamefont{M.~L.} \bibnamefont{Cohen}}, \bibnamefont{and}
  \bibinfo{author}{\bibfnamefont{S.~G.} \bibnamefont{Louie}},
  \bibinfo{journal}{Phys. Rev. B} \textbf{\bibinfo{volume}{50}},
  \bibinfo{pages}{10780} (\bibinfo{year}{1994}),
  \urlprefix\url{http://link.aps.org/doi/10.1103/PhysRevB.50.10780}.

\bibitem[{cds(1999)}]{cdsmefflandoltboernstein}
in \emph{\bibinfo{booktitle}{II-VI and I-VII Compounds; Semimagnetic
  Compounds}}, edited by
  \bibinfo{editor}{\bibfnamefont{O.}~\bibnamefont{Madelung}},
  \bibinfo{editor}{\bibfnamefont{U.}~\bibnamefont{R\"ossler}},
  \bibnamefont{and} \bibinfo{editor}{\bibfnamefont{M.}~\bibnamefont{Schulz}}
  (\bibinfo{publisher}{Springer Berlin Heidelberg}, \bibinfo{year}{1999}), vol.
  \bibinfo{volume}{41B} of \emph{\bibinfo{series}{Landolt-B\"ornstein - Group
  III Condensed Matter}}, pp. \bibinfo{pages}{1--4}, ISBN
  \bibinfo{isbn}{978-3-540-64964-9},
  \urlprefix\url{http://dx.doi.org/10.1007/10681719_527}.

\bibitem[{\citenamefont{Gutowski et~al.}(2009)\citenamefont{Gutowski, Sebald,
  and Voss}}]{Gutowski2009}
\bibinfo{author}{\bibfnamefont{J.}~\bibnamefont{Gutowski}},
  \bibinfo{author}{\bibfnamefont{K.}~\bibnamefont{Sebald}}, \bibnamefont{and}
  \bibinfo{author}{\bibfnamefont{T.}~\bibnamefont{Voss}}, in
  \emph{\bibinfo{booktitle}{Semiconductors}}, edited by
  \bibinfo{editor}{\bibfnamefont{U.}~\bibnamefont{Roessler}}
  (\bibinfo{publisher}{Springer Berlin Heidelberg}, \bibinfo{year}{2009}), vol.
  \bibinfo{volume}{44B} of \emph{\bibinfo{series}{Landolt-B�rnstein - Group
  III Condensed Matter}}, pp. \bibinfo{pages}{40--43}, ISBN
  \bibinfo{isbn}{978-3-540-74391-0},
  \urlprefix\url{http://dx.doi.org/10.1007/978-3-540-74392-7_27}.

\bibitem[{\citenamefont{Hassan}(1993)}]{Hassan199380}
\bibinfo{author}{\bibfnamefont{A.}~\bibnamefont{Hassan}},
  \bibinfo{journal}{Optics Communications} \textbf{\bibinfo{volume}{98}},
  \bibinfo{pages}{80 } (\bibinfo{year}{1993}), ISSN \bibinfo{issn}{0030-4018},
  \urlprefix\url{http://www.sciencedirect.com/science/article/pii/003040189390762T}.

\bibitem[{\citenamefont{H\"{u}gel et~al.}(2000)\citenamefont{H\"{u}gel,
  Heinrich, and Wegener}}]{Hugel2000}
\bibinfo{author}{\bibfnamefont{W.}~\bibnamefont{H\"{u}gel}},
  \bibinfo{author}{\bibfnamefont{M.}~\bibnamefont{Heinrich}}, \bibnamefont{and}
  \bibinfo{author}{\bibfnamefont{M.}~\bibnamefont{Wegener}},
  \bibinfo{journal}{physica status solidi (b)} \textbf{\bibinfo{volume}{473}},
  \bibinfo{pages}{473} (\bibinfo{year}{2000}),
  \urlprefix\url{http://onlinelibrary.wiley.com/doi/10.1002/1521-3951(200009)221:1\%3C473::AID-PSSB473\%3E3.0.CO;2-I/abstract}.

\bibitem[{\citenamefont{Qi et~al.}(1988)\citenamefont{Qi, Shi, Xiong, and
  Xu}}]{Qi1988575}
\bibinfo{author}{\bibfnamefont{J.}~\bibnamefont{Qi}},
  \bibinfo{author}{\bibfnamefont{K.}~\bibnamefont{Shi}},
  \bibinfo{author}{\bibfnamefont{G.}~\bibnamefont{Xiong}}, \bibnamefont{and}
  \bibinfo{author}{\bibfnamefont{X.}~\bibnamefont{Xu}},
  \bibinfo{journal}{Journal of Luminescence} \textbf{\bibinfo{volume}{40-41}},
  \bibinfo{pages}{575} (\bibinfo{year}{1988}), ISSN \bibinfo{issn}{0022-2313},
  \urlprefix\url{http://www.sciencedirect.com/science/article/pii/0022231388903377}.

\bibitem[{\citenamefont{Maslov and
  Ning}(2003)}]{:/content/aip/journal/apl/83/6/10.1063/1.1599037}
\bibinfo{author}{\bibfnamefont{A.~V.} \bibnamefont{Maslov}} \bibnamefont{and}
  \bibinfo{author}{\bibfnamefont{C.~Z.} \bibnamefont{Ning}},
  \bibinfo{journal}{Applied Physics Letters} \textbf{\bibinfo{volume}{83}},
  \bibinfo{pages}{1237} (\bibinfo{year}{2003}),
  \urlprefix\url{http://scitation.aip.org/content/aip/journal/apl/83/6/10.1063/1.1599037}.

\bibitem[{\citenamefont{Maslov and Ning}(2004)}]{1337019}
\bibinfo{author}{\bibfnamefont{A.}~\bibnamefont{Maslov}} \bibnamefont{and}
  \bibinfo{author}{\bibfnamefont{C.-Z.} \bibnamefont{Ning}},
  \bibinfo{journal}{Quantum Electronics, IEEE Journal of}
  \textbf{\bibinfo{volume}{40}}, \bibinfo{pages}{1389} (\bibinfo{year}{2004}),
  ISSN \bibinfo{issn}{0018-9197}.

\bibitem[{\citenamefont{Roeder et~al.}(2014)\citenamefont{Roeder, Ploss,
  Kriesch, Buschlinger, Geburt, Peschel, and Ronning}}]{Roeder2014}
\bibinfo{author}{\bibfnamefont{R.}~\bibnamefont{Roeder}},
  \bibinfo{author}{\bibfnamefont{D.}~\bibnamefont{Ploss}},
  \bibinfo{author}{\bibfnamefont{A.}~\bibnamefont{Kriesch}},
  \bibinfo{author}{\bibfnamefont{R.}~\bibnamefont{Buschlinger}},
  \bibinfo{author}{\bibfnamefont{S.}~\bibnamefont{Geburt}},
  \bibinfo{author}{\bibfnamefont{U.}~\bibnamefont{Peschel}}, \bibnamefont{and}
  \bibinfo{author}{\bibfnamefont{C.}~\bibnamefont{Ronning}},
  \bibinfo{journal}{Journal of Physics D} \textbf{\bibinfo{volume}{47}},
  \bibinfo{pages}{394012} (\bibinfo{year}{2014}),
  \urlprefix\url{http://dx.doi.org/10.1088/0022-3727/47/39/394012
  http://arxiv.org/abs/1407.6744}.

\end{thebibliography}
\end{document}